\documentclass[aps,nofootinbib,superscriptaddress,twocolumn,preprintnumbers]{revtex4}
\usepackage{graphics,amsmath,amssymb,amsbsy,latexsym,verbatim}
\usepackage{hyperref}

\usepackage{setspace}
\usepackage{amsmath,amssymb,amsfonts,amsthm,bbm}
\usepackage{graphicx}
\usepackage{enumerate}

\newcommand{\be}{\nopagebreak[3]\begin{equation}}
\newcommand{\ee}{\end{equation}}
\newcommand{\ba}{\nopagebreak[3]\begin{eqnarray}}
\newcommand{\ea}{\end{eqnarray}}

\newcommand{\bc}{}

\newcommand{\Ref}[1]{(\ref{#1})}

\begin{document}
\title{ \Large On the geometry of loop quantum gravity on a graph} 

    \author{Carlo Rovelli and Simone Speziale}
\affiliation{Centre de Physique Th\'eorique,\footnote{Unit\'e Mixte de Recherche (UMR 6207) du CNRS et des Universit\'es Aix-Marseille I, Aix-Marseille II et du Sud Toulon-Var. Laboratoire affili\'e \`a la FRUMAM (FR 2291).} CNRS-Luminy Case 907, 13288 Marseille Cedex 09, France}

\date{\small \today}
\begin{abstract}

\noindent  
We discuss the meaning of geometrical constructions  associated to loop quantum gravity states on a graph. In particular, we discuss the ``twisted geometries" and derive a simple relation between these and Regge geometries.\\

\end{abstract}

\maketitle

\section{Introduction}

The state space of loop quantum gravity (LQG) is the direct sum of Hilbert spaces ${\cal H}_\Gamma$ associated to graphs $\Gamma$ \cite{Rovelli,Thiemann,Ashtekar:2004eh,Rovelli:2010wq}. In practical applications, a convenient approximation is obtained by cutting-off the continuum theory to a single fixed graph.  Can we assign a geometry to the states on a single graph $\Gamma$?  

The idea of relating loop-quantum-gravity states with discrete geometries goes back to Immirzi \cite{Immirzi:1995yk}. Geometric interpretations of this kind have recently been discussed in the literature.  A detailed interpretation in terms of discrete ``twisted geometries" has been proposed in \cite{Freidel:2010aq}.  An interpretation in terms of a mode expansion of the geometry of a 3-sphere has been discussed in \cite{Rovelli:2008dx,Battisti:2009kp}. Similarly, relations between loop gravity (or loop-inspired models) and discrete geometries have been considered in \cite{Dittrich:2008ar,Bonzom:2009wm,Oriti:2009wg}. 
On the other hand, a common point of view  in the literature is that LQG states do not need geometrical interpretations of this kind (see \cite{Rovelli,Thiemann,Thiemann:2002vj,Ashtekar:2004eh}).  

In this note we contribute to the clarification of this issue with two comments.  First, we discuss the general meaning of assigning a specific geometrical interpretation to states in ${\cal H}_\Gamma$, and the compatibility between different choices. 

The second comment regards the geometrical interpretation of the twisted geometries. We show that in the special case in which the dynamical variables are compatible with a Regge geometry, the canonical transformation studied in \cite{Freidel:2010aq}, which defines the ``twisted geometries", is precisely given by the explicit computation of the holonomy and the flux of the electric field over a given 4d Regge geometry. 

\section{Assigning geometries to states}

A basis of LQG states is given by the spin network states. A spin network state has support on a graph $\Gamma$ and determines a 3d ``quantum geometry". The (intrinsic) geometry is discrete, and can be visualized as made up of ``quanta of space", or ``polymeric'', or similar \cite{Rovelli:1989za,Ashtekar:2004eh,Rovelli}. At the same time, the extrinsic curvature is completly fuzzy, due to the Heisenberg principle. To bridge to a semiclassical description of space, we can consider coherent states peaked (but not sharp) on both the intrinsic and the extrinsic geometry \cite{Thiemann:2002vj}. These are labelled by a continuous classical 3d geometry $(E^a_i(x),A^i_a(x))$ and have support on all possible graphs. 

In order to extract physics from theory, on the other hand, we often need to rely on approximations. A convenient one is to allow only states living on a fixed finite graph $\Gamma$. The Hilbert space ${\cal H}_\Gamma$, formed by the states with support on $\Gamma$ (or subgraphs of $\Gamma$), provides a \emph{truncation} of the theory, which may be sufficient to capture the physics of appropriate regimes \cite{Rovelli:2008dx,Rovelli:2010wq}. Now, one can consider coherent  states also in the Hilbert space ${\cal H}_\Gamma$ of the truncated theory. However, in what sense can one assign a classical geometrical interpretation to \emph{these} states?  This is the problem we address here. 

For a given graph $\Gamma$ with $L$ links $l$  and $N$ nodes $n$, ${\cal H}_\Gamma$ is the space of square integrable functions $\psi(U_l)$ over the group manifold $SU(2)^L$, invariant under the gauge transformations $U_l\to V_{s(l)} U_l V^{-1}_{t(l)}$, where $U,V\in SU(2)$ and $s(l)$ and $t(l)$ are the source node and the target node of the link $l$.  The elementary operators defined on  ${\cal H}_\Gamma$ are the multiplicative operators $U_l$ and the right invariant vector fields $X_l$ which are the generators of the left action of $SU(2)$.  Consider the set of the operators $(U_l,X_l)$ defined on ${\cal H}_\Gamma$ (for a fixed graph $\Gamma$). The corresponding classical variables (which we also indicate as $(U_l,X_l)$) parametrize the phase space $T^*SU(2)^L\cong (su(2)\times SU(2))^L)$, thus a coherent state in  ${\cal H}_\Gamma$ will be peaked on a point $(U_l,X_l)$ in this phase space.  Can we interpret a point in this phase space, that is, the set $(U_l,X_l)$, as a (intrinsic and extrinsic) 3d geometry? 

The difficulty is due to the following reason. The operators $(U_l,X_l)$ have a well-defined interpretation: they are (the restriction to ${\cal H}_\Gamma$ of) the holonomy of the Ashtekar-Barbero connection $A$ along the link $l$, and the flux of the Ashtekar's electric field $E$ over any surface that intersects the sole link $l$ of $\Gamma$ in the immediate proximity of its source. They capture only a \emph{finite} number of degrees of freedom, out of the infinite number of the degrees of freedom of the continuous gravitational field.  The value of the observables $U_l$ and $X_l$ on a single graph, therefore, is not sufficient to determine a continuous gravitational field \emph{uniquely}. Therefore, it cannot determine a 3d geometry completely.  In fact, we are here in context of a truncation of the full theory, where the continuous metric is replaced by a finite set of variables.  The set  $(U_l,X_l)$ characterizes a geometry only partially, in the same sense as when we partially characterize a continuous function $f(x)$ by means of a finite number of its values $f_n=f(x_n)$.

However, in physics, when we are given a finite data set $(x_n, f_n)$, it is often convenient to choose an algorithm to select a preferred \emph{interpolating} function $f(x)$, namely a function such that $f_n=f(x_n)$. The interpolation procedure is of course vastly under-determined, but it is nevertheless often convenient to \emph{choose} an interpolating function.  
Several choices are common.  For instance, if we have $N$ points $x_n$ (say in the interval $[0,2\pi]$), we can choose: (i) the interpolating polynomial $f(x)=\sum_{k=1}^N\ a_k x^k$, with coefficients $a_k$ determined by $\sum_{n}a_kx_n^k=f_n$; or (ii) the periodic function  $f(x)=\sum_{k=1}^N e^{ikx}\ c_k$; or (iii) the piecewise linear function that takes the value $f(x)=f_{n}+f_{n+1}(x-x_n)/(x_{n+1}-x_n)$ for $x\in[x_{n+1},x_n]$ ; or (iv) the discontinuous piecewise constant function that takes the values 
\be
f(x)=f_n,\hspace{3em} {\rm for}\ \   n<x<n+1;
\label{interp}
\ee
and so on. Each of this choices has specific advantages, and each is useful in order to visualize the data set.  Can we do something similar with the geometrical data $U_l$ and $X_l$?

\begin{figure}
\centering
\includegraphics[width=0.115\textwidth]{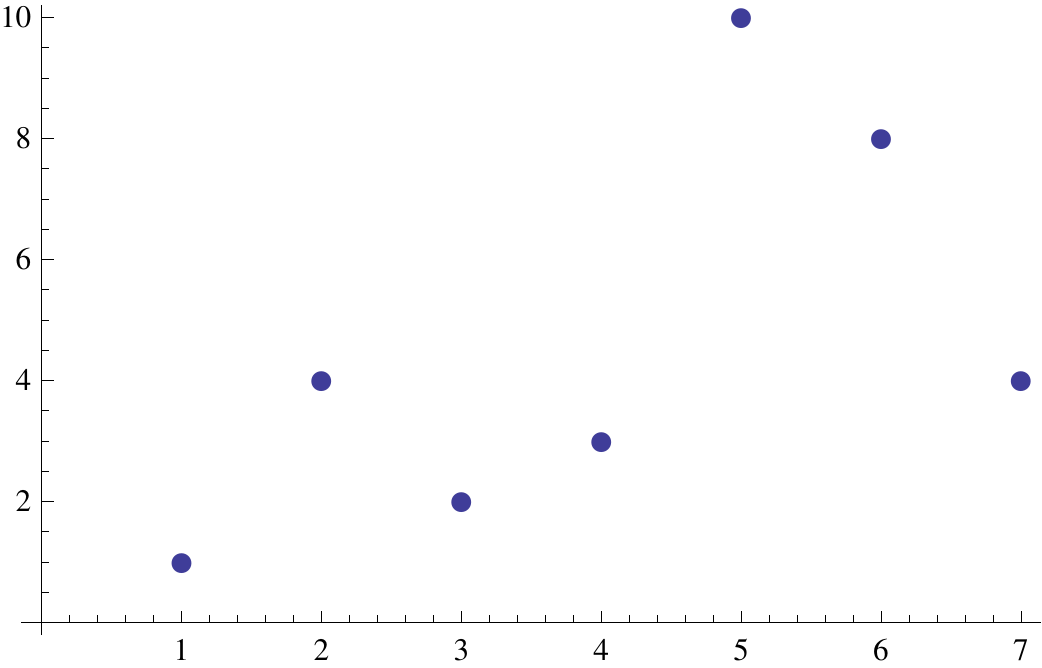}
\includegraphics[width=0.115\textwidth]{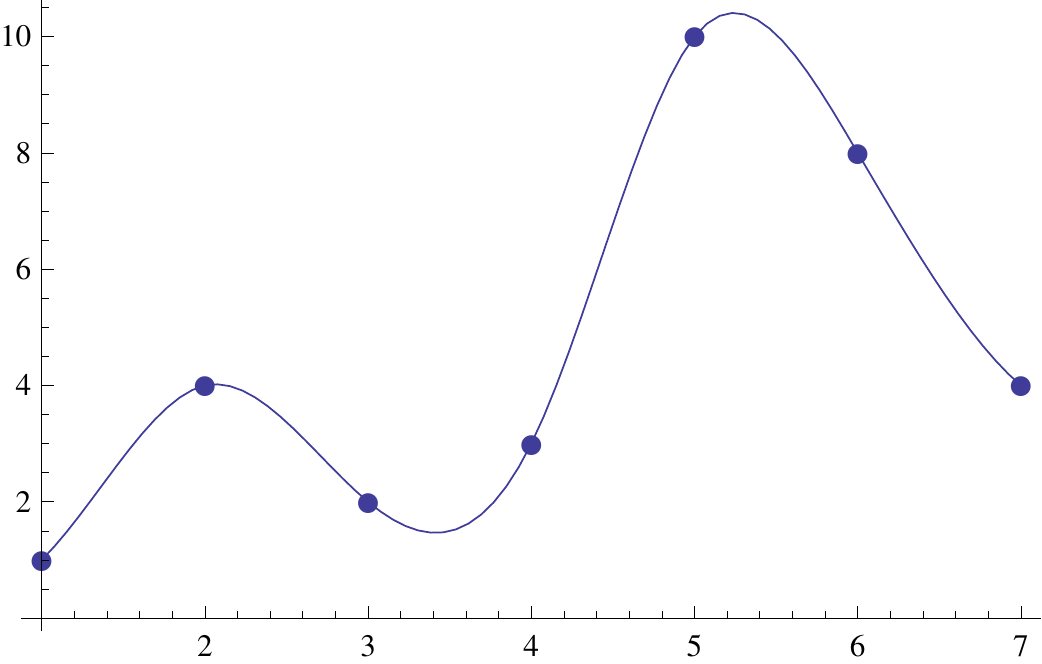}
\includegraphics[width=0.115\textwidth]{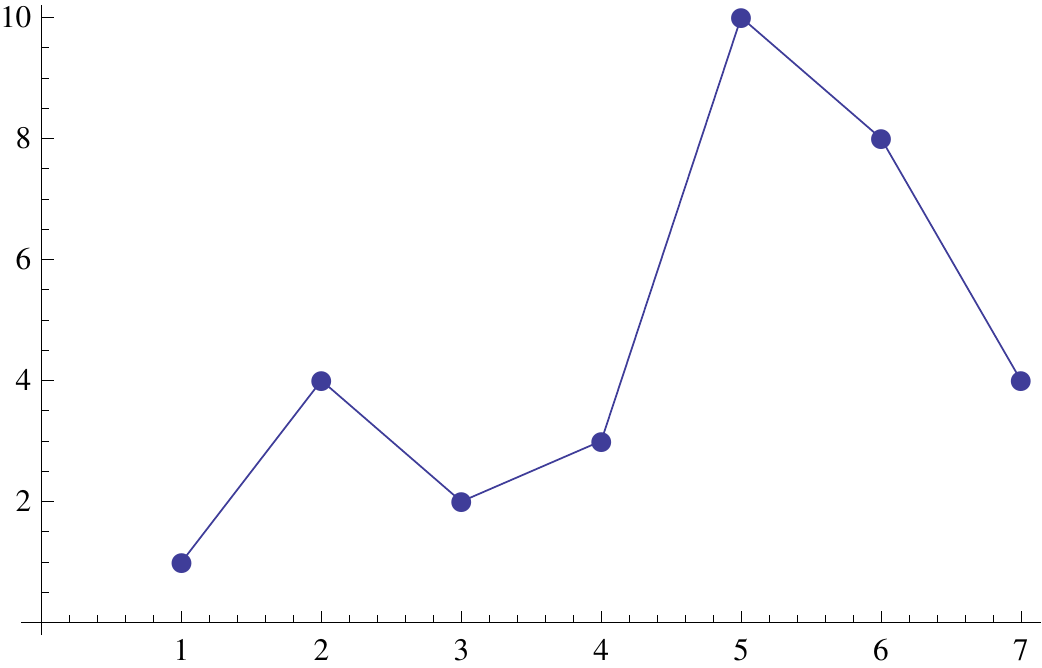}
\includegraphics[width=0.115\textwidth]{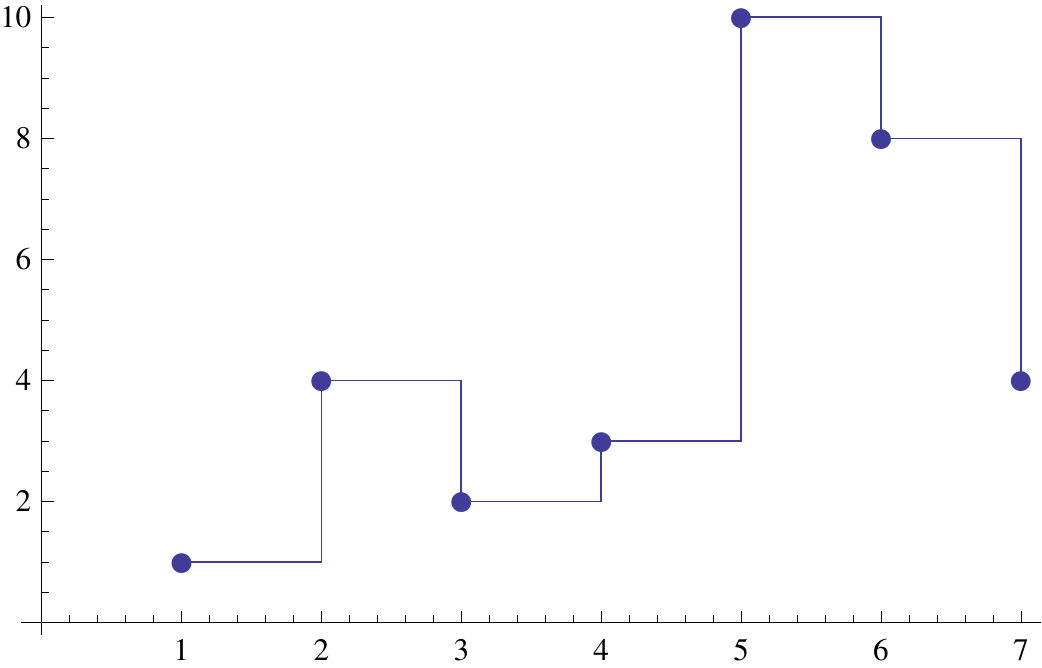}
\caption{A data set and various interpolating functions: polynomial (cfr. mode expansion in cosmology), piecewise linear (cfr. Regge geometries) and piecewise flat (discontinuous, cfr. twisted geometries for generic holonomy-fluxes).}
\label{fig1}
\end{figure}

That is, can we find an algorithm that picks up a preferred ``interpolating geometry" for each set of data $(U_l,X_l)$?   Making this choice is not strictly needed for the interpretation of the theory. LQG is a continuous theory defined by a set of variables much larger than the  $(U_l,X_l)$ of a single fixed graph. However, in the context of a truncation we restrict our attention to a finite number of gravitational field variables, and it is interesting to choose an algorithm that selects a preferred geometry characterized by the data $U_l$ and $X_l$. The algorithm is not unique, but a good choice may serve the purpose of providing a geometrical intuition for the restricted set of gravitational variables $U_l$ and $X_l$.  This is the sense in which a geometry can be associated to the set  $(U_l, X_l)$: 
the interpolating result provides an \emph{approximation} of a continuous geometry.

Two such choices have been recently considered in the literature.  In \cite{Rovelli:2008dx,Battisti:2009kp,Bianchi:2010zs} the idea of a mode expansion of the geometrical degrees of freedom of a (compact) space in hyperspherical harmonics has been put forward in the context of loop quantum cosmology \cite{Ashtekar:2007tv,Bojowald:2008zzb}.  Roughly speaking, it goes as follows. Consider a three-sphere $S^3$ with a smooth metric on it. The components of the 4d metric $g({\alpha}), {\alpha}\in S^3$ can be expanded on a basis of Wigner's $D$ functions,
$$ 
g({\alpha}) = \sum_{j,m,n} g^{jmn} D^{(j)}_{mn}({\alpha}).
$$
If we truncate the expansion to a finite order in $j$, we obtain a set of 3d geometries where ``short wavelength" modes are not excited. We can compute the quantities  $U_l$ and $X_l$ for a finite graph (and its dual cellular complex) on the resulting geometries;  these are then expressed as functions of the modes amplitudes $g^{jmn}$ \cite{Battisti:2009kp}. Solving for $g^{jmn}$ we obtain a geometry for each set $(U_l, X_l)$. Notice that this is the analog of using the data set $f_n$ for fixing the amplitudes of the first $N$ Fourier components of $f(x)$, namely the analog of the example (ii) mentioned above.

Alternatively, one may capture this finite amount of information with a \emph{discrete} metric space. This is an alternative interpolation procedure, analog to the examples (iii) or (iv) above.  A construction of a discrete geometry determined by the variables $(U_l, X_l)$ is discussed in \cite{Freidel:2010aq}.  
The idea  is to introduce a class of discrete metric spaces, called ``twisted geometries", defined over a cellular complex. The geometry is specified by the the set  of variables 
\be\label{defP}
(N_l,\tilde N_l,j_l,\xi_l)\in P_l \equiv S_2\times S_2\times {\mathbbm R} \times S_1
\ee
 associated to each oriented face $l$ of the complex. Each three-cell is taken to be flat, and equipped with an (arbitrary) orthonormal reference system.  The two quantities $N_l$ and $\tilde N_l$ are interpreted as the two (normalized) normals to the face $l$, in the two reference frames associated to the two cells bounded by $l$. The quantity $|j|$ is the area of the face $l$ and the quantity $\xi$ is related to the extrinsic curvature of the complex at $l$ (in a manner that we clarify in this paper).  The relation between the variables  $(N_l,\tilde N_l,j_l,\xi_l)$ that specify a twisted geometry and the LQG variables is given by the canonical transformation (dropping the suffix ${}_l$) 
\be
X=jn\tau_3\tilde n^{-1},\hspace{3em}  U=n e^{\xi\tau_3}\tilde n^{-1}.
\label{pippo}
\ee
Here $X=X^i\tau_i\in su(2)$ where $\tau_i, i=1,2,3$ are the Pauli matrices multiplied by $-i/2$, and $n=n(N)\in SU(2)$ is defined by 
\be
N^i=R^i{}_j(n)z^j.
\label{phase}
\ee
where $\vec z=(z^j)=(0,0,1)$, $R^i{}_j(n)$ is the adjoint representation of $SU(2)$ (equivalently $N^i\tau_i=n \tau_3 n^{-1}$), and by (the ``phase convention")
\be
R^i{}_j(n)(\vec z\times\vec N)^j=(\vec z\times\vec N)^i.
\ee
From the quantities above one can construct a metric, but it is in general discontinuous on the faces. In particular,  the limit of the length of a line that approaches a 1-cell depends on the side from which the 1-cell is approached. This discontinuity is analogous to the discontinuity in \eqref{interp}.

In the next section we illustrate in some detail this discontinuity and the metric structure of such twisted geometries. In particular, we show that a Regge geometry is a special case of twisted geometry, and we give a simple derivation of \eqref{pippo} in this case.  This will show in particular that $\xi$ can be directly obtained from the angle between the 4d normals to the 3-cells.

\section{Twisted geometries from Regge geometries}

Consider a four-dimensional Regge manifold.  By this we mean a metric space composed by glued flat four-simplices, with matching geometry at the intersections. Consider a three dimensional slice $\Sigma$ in this manifold. The slice is formed by a collection of tetrahedra and has intrinsic and extrinsic curvature.  The intrinsic curvature in concentrated on the edges, while the extrinsic curvature is concentrated on the triangles. Let $q_{ab}(x)$ be the metric tensor, in some coordinate system.   Let $e_a(x)=e^i_a(x)\tau_i$ be a corresponding co-triad, in some gauge.  These can be taken to be continuous, but, in general, not necessarily smooth, because of the distributional curvature on the edges. Thus, the metric is continuous but non differentiable, like in the interpolating function of the example (iii) above. 

Let $k_{ab}(x)$ be the extrinsic curvature. Notice that on a Regge manifold the extrinsic curvature vanishes inside each tetrahedron, since each tetrahedron is flat, and is concentrated on the triangles. Consider in particular a triangle $f$ and a cartesian coordinate system that covers the triangle (this is always possible in a Regge manifold, of course), and let $N_a$ be the 3d normal to this triangle in this coordinate system (we use a capital letter instead of the more common low-case notation for the normal, to avoid confusion with the $n\in su(2)$ group elements considered above).  In these coordinates $k_{ab}(x)$ is constant along the triangle, that is
\be
 k_{ab}(x)= k_{ab} \int_f \delta^3(x,f(\sigma))\, d^2\sigma.
 \label{ec}
\ee 

An important observation is that the matrix $k_{ab}$ has a particular form. 
First, the extrinsic curvature is the derivative of the 4d normal to $\Sigma$. Since the normal changes only across $f$, the derivative is non vanishing only in the direction $N_a$ normal to $f$. Second, in a Regge geometry,  when moving across the surface,  the normal rotates in a plane normal to $f$. This is because the normals of both tetrahedra are orthogonal to the triangle; hence the difference is also orthogonal to the triangle.  Therefore \emph{both} indices of $k_{ab}$ are nonvanishing only in the direction of the normal to the triangle
\be
 k_{ab}= k\  N_a N_b,
 \label{ec2}
\ee 
The value of $k$ is then simply the curvature of a curve at a point where there is an angle $\theta$, where $\theta$ is the dihedral angle between the 4d normals to the two tetrahedra at $f$.  Such curvature can be computed by approximating the angle with an arc of angle $\theta$ in a circle of radius $\epsilon$, and hence curvature $1/\epsilon$. The integral of the curvature along the curve is
\be
\int k(s)ds=\int_0^{\epsilon\theta}  \frac1\epsilon\ ds=\theta
\ee
Therefore, in the limit $\epsilon\to0$ we have $k(s)=\theta\delta(s)$. Comparing with \eqref{ec} and \eqref{ec2}, we have finally 
\be
 k = \theta.
\ee 

Consider now a graph dual to the triangulation $\Sigma$. That is, a graph with a 4-valent node inside each tetrahedron and a link $l$ crossing each triangle $f$. Consider the holonomy-flux variables $U_l$ and $X_l$, where $l$ in $U_l$ is a link of the graph and $l$ in $X_l$ is the corresponding dual triangle $f$. These can be explicitly \emph{computed} on the given Regge geometry, according to their definition
\be
 X^i_l=\int_f E^{i}=\int_f E^{ai}N_a\; d^2\sigma
\ee
in the frame of the source of $l$, and
\be
 U_l={\cal P}\exp{\int_l A}={\cal P}\exp{\int_l dl^a(\Gamma_a^i+\gamma e^{bi}k_{ab})\tau_i}. 
\ee
Here $E^{ai}$ is the Ashtekar electric field, namely the inverse densitized triad, 
$A_a^i=\Gamma_a^i+\gamma k_a^i$ is the Ashtekar-Barbero connection, where $\Gamma_a^i$ is the 3d spin connection, $\gamma$ the Immirzi parameter, and $ k_a^i=e^{bi}k_{ab}$ with $e^{bi}$ is the triad field. Finally, 
$\cal P$ is the path ordering along $l$,  and $dl^a$ is the line element along to the link $l$. 

Let us evaluate these variables on $\Sigma$. It is convenient to start by choosing a gauge for $e^i_a(x)$.  It will then be easy to transform the variables to arbitrary gauges. Given a tetrahedron $t$, fix cartesian coordinates that cover the tetrahedron, its four faces, as well as the entirety of the four edges dual to the four faces.  Then $q_{ab}(x)= \delta_{ab}$ and we can choose a gauge where $e_{a}^i(x)= \delta_{a}^i$ on this coordinate patch.  Then we have immediately
\be\label{X2}
 X^i=\int_f E^{ai}N_a d^2\sigma = \delta^{ai}N_a \int_f d^2\sigma =  j N^i
\ee
where $j$ is the area of the face and $N^i=e^{ia}N_a$ is the normal to the face in the coordinate system chosen. Next, consider two adjacent tetrahedra. We can extend the cartesian coordinate system to the second tetrahedron. Because of the 3d flatness, the spin connection part of the connection vanishes, and we are left with 
\be
 U_l={\cal P}\exp{\gamma \int_l dl^a \delta^{ib}k_{ab}\tau_i}
 \ee
Inserting the explicit form \eqref{ec} and  \eqref{ec2}  of the extrinsic curvature into this equation we have
\be
U_l={\cal P}\exp{\gamma\theta\; \delta^{bi}N_b \tau_i\int_l\int_f \delta^3(l,f(\sigma)) dl^a  N_ad^2\sigma}.
\ee
But the integration is precisely the definition of the intersection number, which is unit. Hence we have simply
\be
U_l={\cal P}\exp{\gamma\theta N^i\tau_i} .\label{U1}
 \ee
 
So far we have worked in a gauge in which the two tetrahedra adjacent to the face share the  same reference frame. Let us now rotate the second of these with an arbitrary $SO(3)$ rotation. Then the parallel transport $U_l$ gets an additional contribution 
$U_l\to U_l V$ where $V \in SU(2)$ is the rotation that rotates the first reference frame into the second.  Let us parametrize $V$ with a unit vector $\tilde N$ and an angle $\alpha$. We then write $V \equiv n\tilde n^{-1}_\alpha$, where $n=n(N)\in SU(2)$ is defined by \eqref{phase}, and analogously $\tilde n_\alpha = \tilde n(\tilde N) e^{\alpha\tau_3} $.
Multiplying \Ref{U1} from the right by $V$ we get
\be\label{U2}
U_l=n \; e^{(\gamma\theta-\alpha)\tau_3}\; \tilde n^{-1}.
\ee

The expressions \Ref{X2} and \Ref{U2} reproduce precisely the canonical transformation \eqref{pippo} considered in \cite{Freidel:2010aq}, which map the holonomy-flux variables $(U,X)_l$ into \emph{twisted geometry} variables $(j,\xi,N,\tilde N)_l$, where we recognize $\xi$ as
\be
      \xi= \gamma\, \theta - \alpha,
\ee
namely (up to gauge and the Immirzi parameter) as the modulus of the extrinsic curvature, that is the dihedral angle between the 4d normals of a Regge geometry.\footnote{Notice that a similar relation arises in the semiclassical limit of the new spin foam models, e.g. \cite{Barrett:2009cj}.}

This calculation gives a simple geometrical interpretation to \eqref{pippo}, in the sense that it shows that whenever a Regge geometry is available, the variables of twisted geometries coincide precisely with the \emph{evaluation} of the holonomy-flux variables on this Regge geometry. The underlying Regge geometry can therefore be chosen as the interpolating geometry for these particular $(U_l, X_l)$ configurations. 

\section{Twisted geometries that are not Regge geometries}

Holonomies and fluxes computed as above from the Regge geometries automatically satisfy certain conditions, namely that the length of the edges of a triangle is the same when computed in the frames of the two tetrahedra sharing it. This means that the  metric induced on a given triangle is continuous. However, these gluing or ``shape matching'' conditions \cite{Dittrich:2008va} are not satisfied by a generic point $(U_l,X_l)$ in the (gauge-invariant) phase space of loop gravity \cite{Dittrich:2008ar,Freidel:2010aq,Bahr:2009ku}. As a consequence, an interpolating geometry in terms of piecewise flat continuous metrics is not possible in general, but only on the (measure zero) subspace  where the shape matching conditions hold. In a sense, Regge geometries are ``too rigid'' to be able to interpolate an arbitrary holonomy-flux configuration.

What can the interpolating geometries be then, in the general case? The above discussion suggests that we could insist on piecewise-flat discrete geometries, upon giving up the continuity. 
In fact, the result of \cite{Freidel:2010aq} is that the full (gauge-invariant) phase space can be still described by the variables $(j, \xi, N, \tilde N)_l$, except they now have a larger range: instead of being only the ones coming from a Regge geometry as above, they now span the space 
\be\label{P0}
P^0 \equiv \times_l\  P_l\; /\!/\; C,
\ee 
where $C$ is the closure condition 
\be
C_n=\sum_{l\in n}j_l N_l=0
\label{gauge}
\ee 
at each node $n$; and $/\!/$ indicates the restriction to $C=0$ and the factorization by the orbits generated by $C$. 
Thanks to the closure conditions, the set of the variables $(N_l, j_l)$ for the links adjacent to each given 4-valent node $n$ can still be interpreted as determining the local geometry of a flat tetrahedron.  This fixes the metric inside each tetrahedron. 
However, while the area of the face is the same when measured as a limit coming from one side or the other, nothing now guarantees that the length of the edges match.
The difference with the Regge case is that the metric is in general \emph{discontinuous} across the triangles. 
Thus, the interpolating geometry is obtained gluing flat tetrahedra in such a way that the metric across the faces is discontinuous.    This is analogous to the value of the interpolating function \eqref{interp} above, where the value of the function of the integers is different if defined as a limit from one side or the other.  
These are the (gauge-invariant, or \emph{closed}) twisted geometries \Ref{P0}: a particular choice of interpolating geometry, which \emph{is valid for any point in the phase space of loop gravity}, and which reduces to a Regge geometry when the shape matching conditions are satisfied.

An advantage of this construction is that it extends to nodes of arbitrary valence, and thus arbitrary graphs not only dual to triangulations. In this case, the twisted geometry is assigned to a cellular decomposition dual to the graph, in which each node is dual to a \emph{polyhedron}, and each link to a polygon \cite{Conrady:2009px,Freidel:2009ck,Dona}.

\vspace{1.5em}

The twisted geometry parameterization extends to the non-gauge-invariant level. The kinematical phase space is given by $\times_l P_l \cong T^*SU(2){}^L$ (see \cite{Freidel:2010aq} for details), with \eqref{gauge} relaxed. To each link is thus assigned an angle, the (oriented) area of the face dual to it, and the two normals as seen from the two frames sharing it. 
Because of the lack of closure conditions, these kinematical twisted geometries do not define a piecewise flat metric even locally.
In this case, the area of a face is still the same when approached from one side or the other.

Finally, it is interesting to consider relaxing also the area matching condition  \cite{noi2}. This leads to a even larger space, in which each link is equipped with the normals, but also \emph{two areas} and two additional angles: $(N,\tilde N, j,\tilde \jmath, \xi,\tilde \xi)$. Remarkably, this eight-dimensional link phase space turns out to span precisely the twistor space ${\mathbbm C}^4$ with canonical Poisson brackets \cite{noi2}. Although this takes us out of the LQG phase space, it is compelling to have such a simple starting point for describing quantum geometry. 

The relations among the different spaces considered are summarized in Table 1.

\begin{table}[ht]
{\framebox{
\begin{tabular}{lclclc|}
Twistor space & & \\ 
& & \\
\multicolumn{3}{l}{\hspace{.7cm} $\downarrow$ \emph{area matching reduction}} \\
& & \\
Twisted geometries & $\Longleftrightarrow$ &\ \ phase space of  \\
\ & & \ \ loop gravity  \\
\hspace{.7cm} $\downarrow$ \emph{closure reduction} & \\
& & \\
Closed twisted geometries  & 
$\Longleftrightarrow$ & \ \ gauge-invariant \\
& & \ \ loop gravity \\
\multicolumn{3}{l}{\hspace{.7cm} $\downarrow$ \emph{shape matching reduction}} \\
& & \\
Regge phase space 
\end{tabular}
}\caption{The relation between the spaces considered for the discrete variables on a truncation of general relativity.}}
\end{table}

\section{Conclusions}

Loop gravity on a fixed graph describes a truncation of general relativity \cite{Rovelli:2010wq}. The variables in this truncation capture only a finite number of the degrees of freedom of the metric. Therefore there is no unique geometric interpretation associated to a single graph. 

``Interpolating" geometries --such as the twisted geometries, multipolar expansions and Regge geometry, discussed here-- are not strictly needed for the physical interpretation of the theory, but provide useful approximations of a continuous geometry. They have important applications, for instance in cosmology, in the study of semiclassical limit in spinfoams \cite{Barrett:2009cj} in the definition of $n$-point functions \cite{Rovelli:2005yj,Speziale:2008uw,Bianchi:2009ri} and in the interpretation of coherent states \cite{Bianchi:2009ky,Freidel:2010}. 

A twisted geometry is a specific choice of ``interpolating geometry", chosen among discontinuous metrics.  To any graph and any holonomy-flux configuration, we can associate a twisted geometry:  a discrete discontinuous geometry on a cellular decomposition space into polyhedra. Thanks to this result, the phase space of LQG on a graph can be visualized not only in terms of holonomies and fluxes, but also in terms of a simple geometrical picture of adjacent flat polyhedra.

We have shown here that in the special case when the holonomy-flux variables admit a Regge interpretation, the canonical trasformation that defines the twisted geometry variables is  precisely given by the explicit computation of the holonomy and the flux of the electric field over the underlying Regge geometry. 

The relation between twisted geometry and Regge calculus implies that holonomies and fluxes carry \emph{more} information than the phase space of Regge calculus.  This is not in contradiction with the fact that the Regge variables and the LQG variables on a fixed graph both provide a truncation of general relativity: simply, they define two distinct truncations of the full theory. 

\subsection*{Acknowledgements}
We are grateful to Eugenio Bianchi for useful discussions.

\vfill

\providecommand{\href}[2]{#2}\begingroup\raggedright\endgroup


\end{document}